\documentclass[11pt,a4paper]{article}
\pdfoutput=1 

\usepackage[font=small,labelfont=bf,
   justification=justified,
  format=plain]{caption}
\usepackage{soul}

\usepackage{jheppub}
\usepackage{latexsym,epsfig,amssymb, amsmath,nicefrac}

\usepackage{bm}
\usepackage{natbib}
\usepackage{graphicx}
\expandafter\let\csname equation*\endcsname\relax
\expandafter\let\csname endequation*\endcsname\relax
\usepackage{amsmath}
\usepackage[labelfont={bf,scriptsize},textfont={scriptsize},justification=RaggedRight,format=hang]{caption,subfig}
\usepackage{color}
\usepackage{blkarray}
\usepackage{xcolor}

\usepackage{subfig} 

\RequirePackage{ifpdf}
\ifpdf
\else 
\fi

\newcommand{\Mp}{M_{\mathrm{P}}}
\newcommand{\cO}{{\cal O}}
\newcommand{\cL}{{\cal L}}
\newcommand{\cN}{{\cal N}}

\newcommand{\vol}{\mathcal{V}}

\newcommand{\beq}{\begin{equation}}
\newcommand{\eeq}{\end{equation}}

\newcommand{\bea}{\begin{eqnarray}}
\newcommand{\eea}{\end{eqnarray}}

\begin{document}

\begin{flushleft}
DESY 18-063
\end{flushleft}

\title{Pole N-flation}

\author[a]{Mafalda Dias,}
\emailAdd{mafalda.dias@desy.de}
\author[a]{Jonathan Frazer,}
\emailAdd{jonathangfrazer@gmail.com}
\author[a]{Ander Retolaza,}
\emailAdd{ander.retolaza@desy.de}
\author[a,b]{Marco Scalisi,}
\emailAdd{marco.scalisi@kuleuven.be}
\author[a]{Alexander Westphal}
\emailAdd{alexander.westphal@desy.de}

\affiliation[a]{Deutsches Elektronen-Synchrotron DESY, Notkestra{\ss}e 85, 22607 Hamburg, Germany}
\affiliation[b]{Institute for Theoretical Physics, KU Leuven, Celestijnenlaan 200D, B-3001 Leuven, Belgium}

\abstract{A second order pole in the scalar kinetic term can lead to a class of inflation models with universal predictions referred to as pole inflation or $\alpha$-attractors. While this kinetic structure is ubiquitous in supergravity effective field theories, realising a consistent UV complete model in e.g. string theory is a non-trivial task. For one, one expects quantum corrections arising in the vicinity of the pole which may spoil the typical attractor dynamics. As a conservative estimate of the range of validity of supergravity models of pole inflation we employ the weak gravity conjecture (WGC). We find that this constrains the accessible part of the inflationary plateau by limiting the decay constant of the axion partner.  
For the original single complex field models, the WGC does not even allow the inflaton to reach the inflationary plateau region.
We analyze if evoking the assistance of $N$ scalar fields from the open string moduli helps addressing these problems. Pole $N$-flation could improve radiative control by reducing the required range of each individual field. However, the WGC bound prohibiting pole inflation for a single such field  persists even for a collective motion of $N$ such scalars if we impose the sublattice WGC. Finally, we outline steps towards an embedding of pole N-flation in type IIB string theory on fibred Calabi-Yau manifolds.

}

\maketitle

\section{Introduction}\label{sec:intro}

The paradigm of cosmological inflation seemingly explains the origin of spatial homogeneity and isotropy, as well as the seeding process for cosmic structure formation. However, its physical origin remains unclear.
High precision studies of the cosmic microwave background have revealed the primordial curvature perturbation to have extremely simple statistics: gaussian to very high precision and describable by just two numbers, the amplitude and tilt of its power spectrum \cite{Ade:2015lrj}. 
It is therefore important to identify what microphysical  mechanism could be   at the origin of this observation.
One possibility is that there is a symmetry present in the underlying theory which ultimately forbids contributions to the inflationary potential capable of giving rise to features in the observed data, \emph{i.e.} a fundamental reason why the inflation potential must take a simple form. A more recent suggestion is that a simple spectrum could be an emergent property of some type of large $N$ dynamics, be it through a large number of terms contributing to a single scalar field potential \cite{Green:2014xqa,Amin:2015ftc,Amin:2017wvc} or by the interaction of a very large number of scalar fields \cite{Dias:2016slx,Dias:2017gva}. A third possibility is that there is some structure in the underlying theory which makes inflation insensitive to a broad array of microphysical details. This would essentially give rise to a universality class, where a diverse range of models result in the same predictions for observable quantities.

A particularly dramatic example of this third possibility is the universality displayed by the class of models termed `pole inflation' \citep{Galante:2014ifa,Broy:2015qna} (or `$\alpha$-attractors'  for a special subclass thereof \citep{Kallosh:2013yoa,Roest:2015qya}). This class of models is defined by the presence of a pole in the kinetic term, such as 
\beq\label{kineticterm1}
\cL_{kin}= -\frac{3\alpha}{4}\frac{(\partial\tau)^2}{\tau^2} \,,\quad \alpha = \cO(1)\,,
\eeq
which renders the dynamics of inflation insensitive to the details of a generic scalar potential, provided this can be expanded around the pole as\footnote{The invariance of the kinetic term \eqref{kineticterm1} under the inversion symmetry $\tau\rightarrow 1/\tau$ makes this model equivalent to one also with negligible kinetic term and a scalar potential $V=V_0 - a_1/ \tau +\mathcal{O}(1/\tau^2)$ .}
\beq
V=V_0 - a_1 \tau +\mathcal{O}(\tau^2)  \, .
\eeq
These models lead to a universal prediction given by $n_s = 1 - 2 / N_e$ and $r=12 \alpha /N_e^2$, where $N_e$ is the number of e-folds of expansion between the pivot scale leaving the horizon and the end of inflation. These predictions are in remarkable agreement with observations and are fully determined by the order of the pole 
(which sets the deviation of $n_s$ from perfect scale invariance; for general pole of order $p$, $n_s=1-p/N_e$ - see e.g. \cite{Broy:2015qna,Terada:2016nqg}),
its residue (which controls the amplitude of primordial gravity waves), and the value of $N_e$, which depends on details of post-inflationary physics.

The kinetic structure of this class of models is of special interest as they can result from logarithmic K\"ahler potentials in 4D $\cN=1$ supergravity effective scenarios,  which are abundant in the context of string compactifications (a more detailed discussion is given in Sec.~\ref{Sec:KinPol}). Schematically, one can have
\beq\label{eq:kahlerpot}
K = - 3\alpha\ln(T+\bar T)\qquad \text{or}\qquad K = - 3\alpha\ln(1- \Phi\bar\Phi)\,,
\eeq
where the first hermitian function is defined in the `half plane' $T+\bar T >0$, while the domain of the second one is the `unit disk' $\Phi\bar \Phi<1$.\footnote{Note that these two K\"{a}hler potentials are related  via the holomorphic relation $\Phi=\frac{T-1}{T+1}$ and a K\"{a}hler transformation (see e.g.~\cite{Carrasco:2015uma,Yamada:2018nsk}).} The corresponding  component of the K\"ahler   metrics are given by 
\beq\label{eq:kahlermetrics}
K_{T \bar T}= \frac{3\alpha}{ (T+\bar T)^2}\qquad \text{or}\qquad K_{\Phi \bar \Phi}=\frac{3\alpha}{ (1- \Phi\bar\Phi)^2}\,,
\eeq
with subscripts denoting partial derivatives. Just as eq.~\eqref{kineticterm1}, they have a second-order pole in the real part of $T$ and radial direction of $\Phi$, respectively, such that these fields can act as the inflaton. Interestingly in this context, the presence of the kinetic pole has the geometric interpretation of the existence of a boundary in moduli space. Note that, as both $T$ and $\Phi$ are complex, the inflaton always comes together with a partner scalar degree of freedom which we will argue to be axionic.

While this class of models arises in 4D $\cN=1$ supergravity (see also \cite{Covi:2008cn,Roest:2013aoa}), it is important to understand if it corresponds to a low-energy effective description that can consistently be embedded in a quantum theory of gravity, like string theory. Various consistency requirements such as the convergence of the higher instanton corrections, the weak gravity conjecture  or the swampland conjectures~\cite{Banks:2003sx,ArkaniHamed:2006dz,Ooguri:2006in} 
place strong bounds on the nature of effective theories that can be embedded in quantum gravity.

A classic example of the importance of this type of bound is provided by natural inflation~\citep{Freese:1990rb}, where a single axion in a shift-symmetric potential is responsible for the cosmological dynamics.  In the effective field theory description, the axionic shift-symmetry protects the potential from radiative corrections such that large periodicities, \textit{i.e.}~potentials with a decay constant $f \gg M_{\rm P}$, can ensure radiative stability even for field displacements larger than $M_{\rm P}$. However, it is not obvious that such symmetries admit  completions in quantum gravity. Specifically, the weak gravity conjecture typically places a bound on the axion decay constant as $f\lesssim\Mp$. But compatibility with observations suggests the axion in natural inflation to have a decay constant $f \gtrsim 5 M_{\rm P}$.

Proposals to evade this contradiction fall into two types:  either they realize an axionic approximate  shift symmetry
via monodromy  (by coupling to a 4-form field strength)~\cite{Silverstein:2008sg,McAllister:2008hb,Kaloper:2008fb,Kaloper:2011jz}, or they use assistance effects driven by several axions participating during inflation. The second case can arise through the tuned alignment of two axions~\cite{Kim:2004rp}, by arranging a hierarchy of the decay constants~\cite{Berg:2009tg,Tye:2014tja,Ben-Dayan:2014zsa}, or as a generic assistance effect driven by a large number of axions termed `N-flation'~\cite{Liddle:1998jc,Dimopoulos:2005ac,Easther:2005zr,Bachlechner:2014hsa,Bachlechner:2017hsj}. 
In the latter, the total inflaton field range $\Delta\varphi$ arises through the collective displacements of individual axions, each of them satisfying the constraints of the WGC: $\Delta\phi_i \sim f_i \lesssim M_{\rm P}$. In the simplest setup of N-flation with $N$ axions with roughly similar decay constants, it is easy to see that the field displacements are related by  $\Delta\varphi \sim \sqrt{N} \Delta\phi_i $, such that for large enough $N$, super-Planckian inflaton displacements seem to be allowed. 
It turns out that generalizing the WGC bounds to theories with multiple axions is more subtle than simply implementing the bound  $f\lesssim\Mp$ for each individual axion.  Implementing WGC bounds for this case implies using the \emph{convex hull condition}~\cite{Cheung:2014vva,Rudelius:2014wla,Rudelius:2015xta,Brown:2015iha}. 
This condition  
 leads to a bound on the collective axionic motion forbidding the $\sqrt{N}$ enhancement with respect to the single field case. This will be further discussed in section \ref{Sec:WGC}.

The situation for a single field driving pole inflation is morally similar. The universal predictions of this scenario arise when the non-canonical field approaches the kinetic pole, or from a supergravity perspective, the boundary of the moduli space. 
However, on generic grounds, precisely in this regime we expect numerous quantum corrections to grow large, thus potentially leading to a loss of control of the setup. We can argue both in 4D effective field theory~\cite{vonGersdorff:2005bf} and in string theory~\cite{Berg:2005ja,Berg:2007wt,Cicoli:2007xp,Haack:2018ufg} for the appearance of such dangerous terms. In this paper, we explore the possibility of using the collective behaviour of a large number $N$ of moduli-like scalar fields   that could  alleviate some of these problems, in what we call \emph{pole N-flation}. Specifically, we propose a scenario where the approach to the pole is achieved by the assistance effect of many fields, such that 
each   field is individually further away from the boundary by a factor of $\sqrt{N}$. 
This  may  suppress some of the generically expected loop contributions which grow large when individual fields approach the boundary.

As a conservative estimate for the domain of validity of the effective description of both pole inflation and pole N-flation, we make use of the WGC constraints.
In both cases the moduli of the simplest supergravity models    
are associated with axionic partners, allowing us to implement bounds on their periodicities as conditions for the consistency with ultra violet physics. 
We find that imposing WGC bounds on the axionic periodicities directly translates into the impossibility of getting close the boundary, and therefore to a finite inflationary plateau in canonically normalized variables. This has dramatic consequences for the viability of pole inflation in general. 
This bound is very stringent for the case of a single superfield, and one might have hoped for a weakening of the WGC-imposed bound when many disk-variable provide a collective `pole $N$-flation' mode. However, it turns out that  in the many axion generalization of the WGC one finds a conspiracy much like in axion $N$-flation to erase any $N$-enhancement. 

The outline of the paper is the following: in \S\ref{Sec:KinPol} we scan the typical kinetic structures derived from string theory and identify the most natural for a scenario with a pole due the collective behaviour of many fields. In \S\ref{Sec:Uni} we will study a supergravity toy model of pole N-flation, describing the ellipsoid structure of the pole and the subsequent universality behaviour. We use these results in \S\ref{Sec:WGC} to establish contact with the WGC and the swampland conjectures, and derive a bound on the field range in pole N-flation. With the aim of embedding these ideas in string theory, in \S\ref{Sec:FibCY} we develop an explicit scenario based on type IIB string theory on fibered Calabi-Yau manifolds. We draw our conclusions in \S\ref{Sec:Concl}. Throughout the paper, we will work in reduced Planck mass units ($\Mp = 1$).

\section{Kinetic poles in string theory}\label{Sec:KinPol}

In order to search for setups with $N\gg 1$ fields and second order kinetic poles, it is illuminating to analyse the structure of kinetic terms in 4D $\cN=1$ supergravity derived from string theory. String compactifications on Calabi-Yau manifolds generically produce K\"ahler potentials containing both closed and open string moduli, with the exact number of such moduli given by the underlying geometry and the amount of D-branes. We can classify the possible K\"ahler potentials as follows:
\begin{enumerate}

\item In perturbative string theory, in the large volume and large complex structure limit, there are at most 3 volume and 3 complex structure moduli which describe the total Calabi-Yau manifold. Together with the axio-dilaton, these correspond to a maximum of 7 chiral fields, which lead to the tree-level K\"ahler potential
\beq
\label{sumsep}
K = - 3\sum_{i=1}^n \alpha_i \ln\left(T_i+\bar T_i\right)\quad,\, n\leq 7 \quad {\rm and} \quad 1\leq 3\sum_{i=1}^n \alpha_i\leq 7\,,
\eeq
with the parameters $\alpha_i$ depending on the number of fields (see e.g.~\cite{Ferrara:2016fwe} in the context of $\alpha$-attractors).

\item There are open string moduli describing brane positions (e.g. D3-branes), and/or open string matter fields as well as gauge fields. Their number is usually subject to tadpole bounds and can be as large as ${\mathcal O}(10^4)$. They appear as a contribution to the volume moduli K\"ahler with the schematic form
\beq\label{Kvm}
K=-3\alpha\ln\left(T+\bar T - \sum_{i=1}^N a_i\Phi_i\bar\Phi_i\right)\,,
\eeq
where the parameter $\alpha$ depends on the specific configuration of the bulk geometry and the K\"ahler moduli.

\item Finally, the moduli space of Calabi-Yau manifolds contains singular regions, most easily seen as the conifold points of complex structure moduli space. Near such singularities, the corresponding moduli acquire a K\"ahler potential of non-polynomial form inside the primary logarithm. For complex structure moduli near a conifold point this generically implies a K\"ahler potential of the form
\beq
K_{c.s.}=-\ln\left(f(u_i,\bar u_{\bar\imath})- z\bar z\ln z\bar z \right)\,,
\eeq
where $z$ denotes the complex structure moduli parametrizing the vicinity of the conifold singularity, and the $u_i$ denote the remaining other complex structure moduli of the Calabi-Yau. The number of such conifold regions in a given Calabi-Yau can be quite large, easily of the order of a few tens.

\end{enumerate}

From this short list we see that a realization of pole inflation involving assistance effects of a large number of fields cannot arise from the large-volume or large-complex-structure type of closed string moduli described in the first class of the list, as their number is intrinsically limited. 
The study of the third class in the list would require an in-depth analysis of multi-conifold complex structure K\"ahler potentials and their dependence on non-conifold complex structure moduli. This analysis is beyond the scope of the present paper and we leave it for future work. In this paper we therefore explore the second class, where a large number of open string moduli fields with K\"ahler potentials of the form  eq.~\eqref{Kvm} lead to the pole N-flation scenario.

\section{The pole N-flation picture}\label{Sec:Uni}

\subsection{Kinetic structure and universality}

We start our analysis by looking at the K\"ahler potential given by eq.~\eqref{Kvm}. Once the K\"ahler modulus $T$ is stabilized, the relevant  dynamics in the EFT is described by 
\begin{equation}
\label{Kpot}
K=- 3\alpha \ln \left(1- \sum_{i=1}^N A_i \ \Phi_i\bar{\Phi}_i\right)\,,
\end{equation}
where the new coefficients are rescaled by the VEV  of $T$, such that $A_i=a_i/\langle T+\bar T\rangle$. The corresponding kinetic term is given by
\begin{equation}
-\sum_{i,j=1}^N K_{\Phi_i \bar{\Phi}_j}\ \partial\Phi^i\partial\bar{\Phi}^j=-3\alpha\sum_{i,j=1}^N \left[\frac{\ A_i A_j\  \bar{\Phi}_i  \Phi_j }{\left(1- \displaystyle\sum_{k=1}^N A_k \ \Phi_k\bar{\Phi}_k\right)^2}+ \frac{\  A_i\ \delta_{ij}}{1- \displaystyle\sum_{k=1}^N  A_k \ \Phi_k\bar{\Phi}_k}\right]\partial\Phi^i\partial\bar{\Phi}^j\,,
\end{equation}
which has a pole for   $R^2 \equiv \sum_k A_k \ \Phi_k\bar{\Phi}_k=1$, which is the equation of an ellipsoid in field space with $N$ independent radii of length directly related to the brane contributions $A_i$. As can be seen by the form of the denominator, the $N$  fields collectively contribute to reach the  boundary without any $\Phi_k$ reaching the boundary itself.\footnote{This is unlike systems where each field has its own distinct pole as in the case   studied in ref.~\citep{Linde:2016uec}. The corresponding K\"ahler potential would be sum-separable of the form of eq.~\eqref{sumsep}. Note that, in this case, the inflationary predictions are strictly related to the specific direction in field space.}  Therefore, when all fields equally contribute to inflation, the displacement of each individual field will be reduced by a factor of  $\sqrt{N}$. This property  could  protect the model against radiative corrections which grow as the fields individually  approach the pole.  Nevertheless, the increase on the number of fields relevant for the model  also implies a larger amount of corrections that may or may not overcompensate the previous gain. Answering this question requires knowing the shape of the quantum corrections which is beyond the scope of this paper and so, we leave for future work. 

To further understand how the presence of these fields affects the model and its dynamics, we make the following change of variables 
\begin{equation}
\begin{aligned}
&\Phi_i=  \frac{R}{\sqrt{A_i}} \Omega_i(\psi_\beta) e^{i\theta_i}\, \\ 
&\bar \Phi_i	= \frac{R}{\sqrt{A_i}} \Omega_i(\psi_\beta)  e^{-i\theta_i}\,,
\label{angularcoords}
\end{aligned}
\end{equation} 
where $\Omega_i(\psi_\beta)$ is the spherical angular element such that $\sum_i \Omega_i^2(\psi_\beta)=1$. Here the index $\beta= 1, \cdots , N-1$.  As will be discussed in  
\S\ref{Sec:WGC}, the angles $\theta_i$ can be associated with axions and will play a crucial role in our understanding of the UV consistency of the setup.

In these variables the line element becomes
\begin{equation}
 -3\alpha \left[\frac{1}{(1-R^2)^2}\ \partial R\partial R + \frac{R^2}{1-R^2} \sum_{i}(\partial_{\beta}\Omega_i)^2 \, \partial \psi_\beta\partial \psi^\beta + \sum_{ij} G_{ij}\ \partial \theta_i \partial \theta_j \right] \, , \label{KMang1}
\end{equation}
where   $\partial_{\beta}\equiv \partial/\partial\psi_\beta$ and
\begin{equation}
G_{ij}=  \frac{R^4}{(1-R^2)^2} \Omega^2_i \Omega^2_j+ \frac{R^2}{1-R^2} \delta_{ij}\Omega^2_i\,. \label{eq:kinetic}
\end{equation}
In this form, the field-space metric has useful features. First, it  is independent of the axionic variables $\theta_i$. It is also diagonal in the variables $R$ and $\psi_\beta$. The mixed terms associated to $\partial R \partial \psi_\beta$ vanish due to $\sum_i \Omega_i\partial_\beta \Omega_i= 1/2\ \partial_\beta \sum_i \Omega^2_i =0$, and so do the terms    $\partial \psi_\beta \partial \psi_\gamma$ for $\beta \neq \gamma$ due to trigonometric relation $\sum_i \partial_\beta\Omega_i\partial_\gamma \Omega_i=0$,  proved in Appendix~\ref{angles}.

\begin{figure}
\captionsetup[subfloat]{farskip=-5pt, captionskip=0pt}
\centering
\vspace{-12mm}
\subfloat{
  \includegraphics[width=78mm]{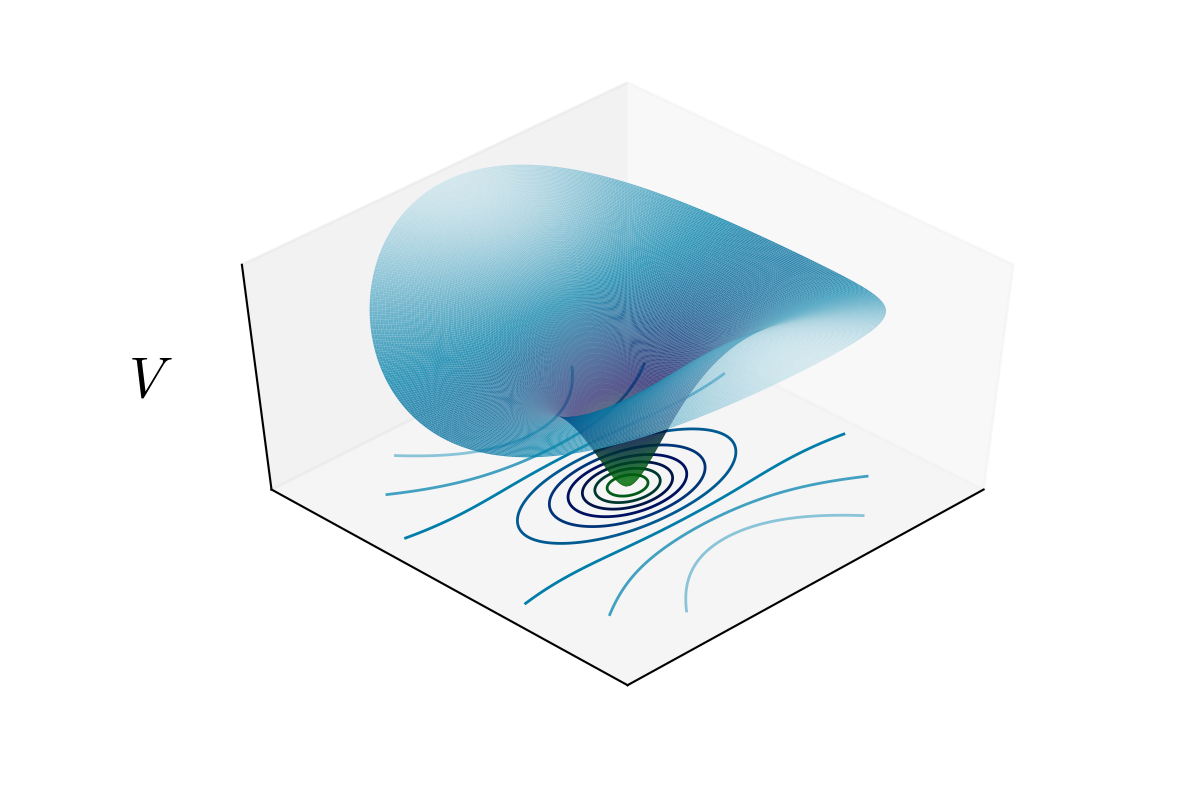}
}
\hspace{-12mm}
\subfloat{
  \includegraphics[width=78mm]{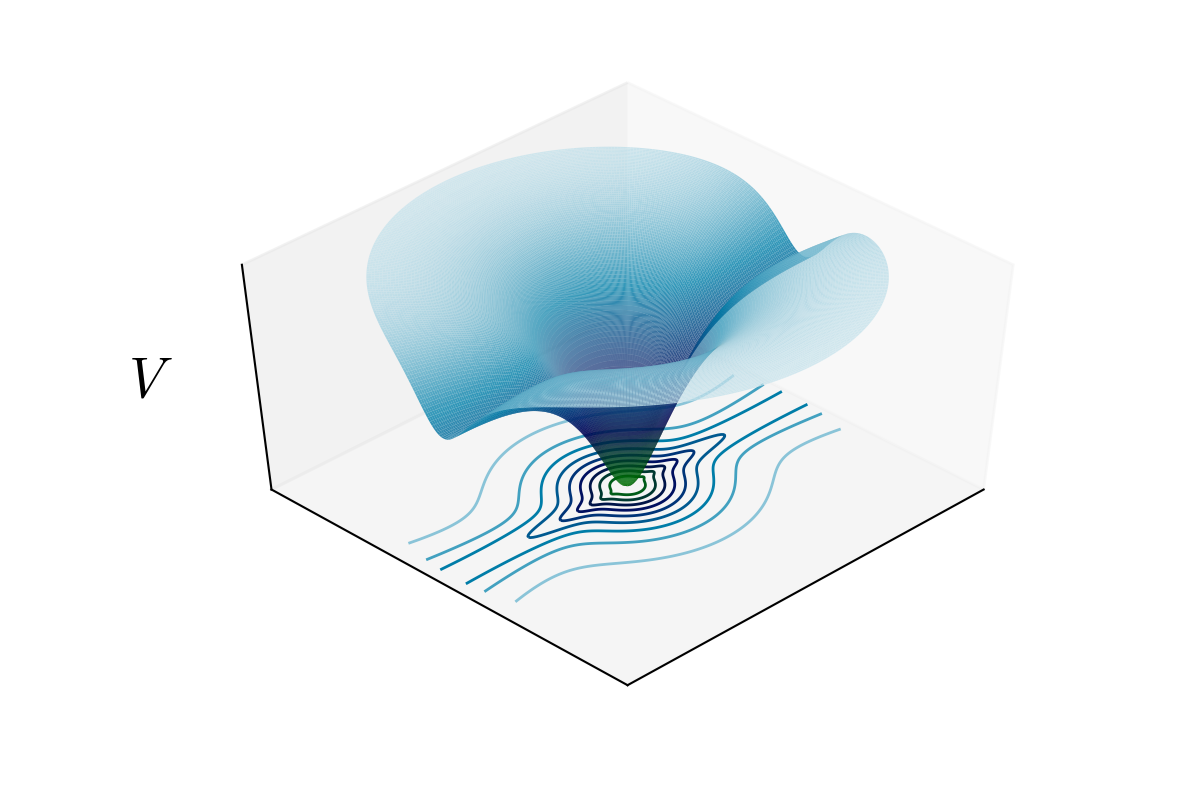}
}
\hspace{+15mm}
\subfloat{
  \includegraphics[width=78mm]{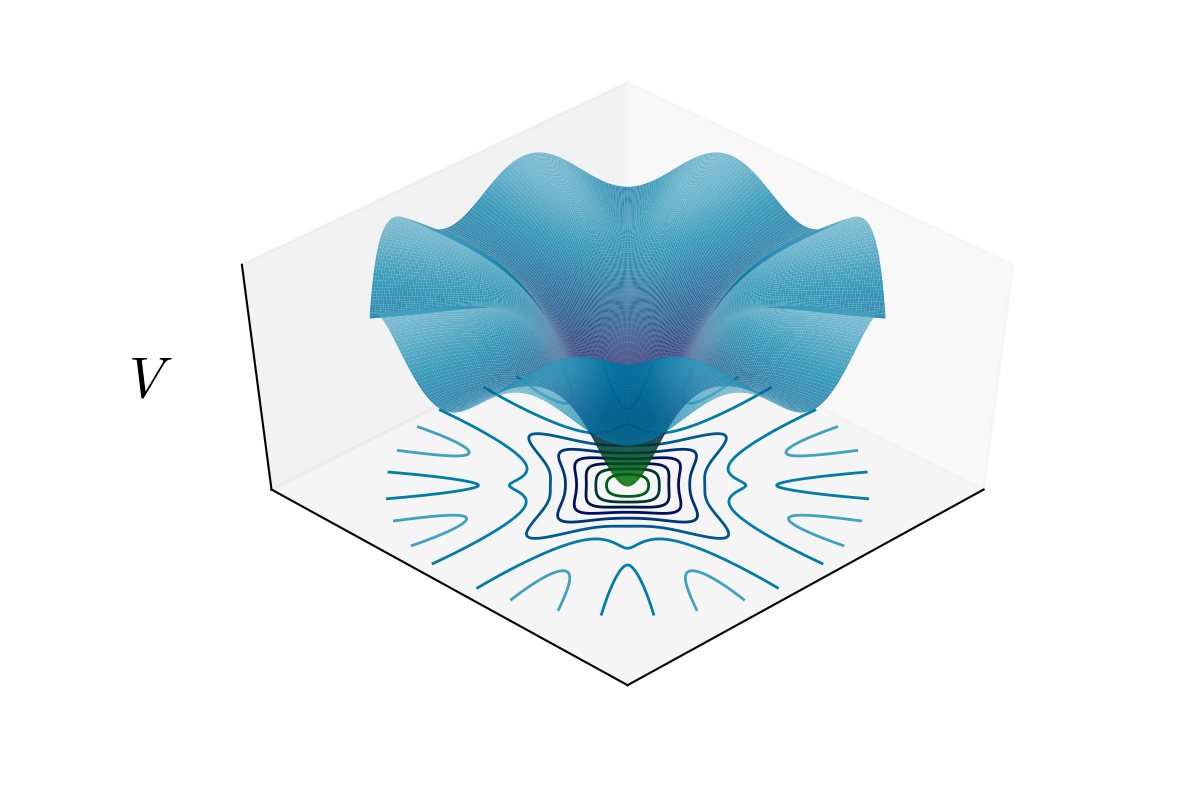}
}
\hspace{-12mm}
\subfloat{
  \includegraphics[width=78mm]{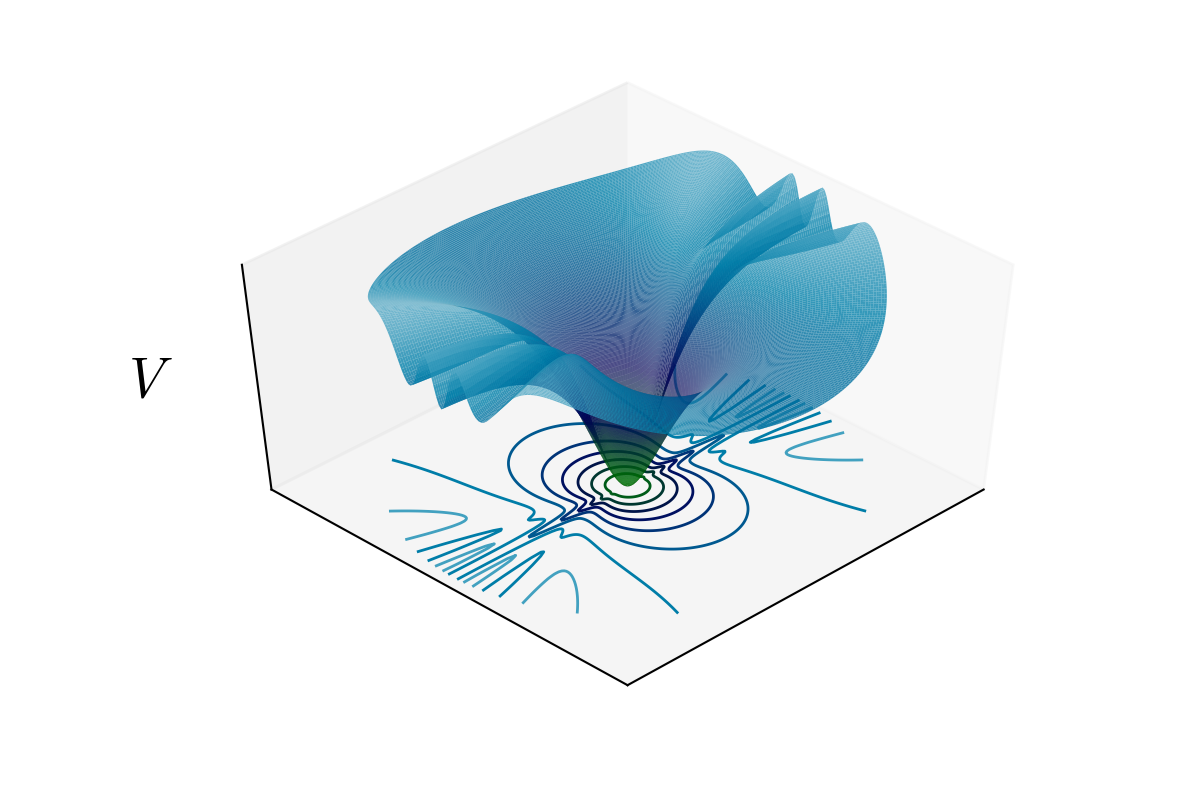}
}
\caption{Potential for the $N=2$ case plotted in the polar coordinates $\{\varphi, \psi\}$, with angles $\theta_1$ and $\theta_2$ minimized. The top-left plot shows the potential eq.~\eqref{VExpansion} up to terms with $p=1$, in the case of equal $A_i$. The top-right plot shows the same potential but for $A_1=1$ and $A_2=0.2$. The bottom-left plot shows the potential with terms up to $p=4$, in the case of equal  $A_i$. The bottom-right plot shows the same potential but for    $A_1=1$ and $A_2=0.2$. We can see that the elliptical structure of the model makes valleys in the potential bundle-up. While different radial directions have plateaus with different amplitudes, the exponential fall-off has the same signature for all initial conditions, or in other words, values of the angular coordinate $\psi_\beta$. The oscillating plateau of the circular case were already noted for a real 2-disk $\alpha$-attractor (without supergravity) in~\cite{Kallosh:2013daa}.}
\label{plots}
\end{figure}

One can easily identify $R$ as the variable with a kinetic pole of second order. Upon canonically normalizing the kinetic term, one has
\begin{equation}
R=\tanh\frac{\varphi}{\sqrt{6\alpha}}\,,
\end{equation}
such that the boundary at $R\rightarrow 1$ is equivalent to $\varphi\rightarrow\infty$. Writing the system in this canonical variable, just like in the single-field pole inflation case, makes evident how the model is stable with respect to considerable deformations of the inflaton scalar potential.  The potential can be generated by means of several mechanisms: via an inflaton-dependent superpotential (with stabilizer superfield~\citep{Kallosh:2013yoa,Scalisi:2015qga} or without it~\citep{Roest:2015qya}), by K\"ahler \cite{McDonough:2016der,Kallosh:2017wnt}, loop~\cite{vonGersdorff:2005bf,Berg:2005ja,Berg:2007wt,Cicoli:2007xp} or higher-derivative~\cite{Ciupke:2015msa} corrections. A generic expansion looks like
\begin{equation}
V=\sum\limits_{ij,p\geq 1}b_{ij,p}( \Phi_i\bar{\Phi}_j + c.c.)^p = \sum\limits_{ij,p\geq 1} b_{ij,p}\ \left[2\frac{\Omega_i\Omega_j}{\sqrt{A_iA_j}}\cos(\theta_i-\theta_j)\ R^{2}\right]^p\,,  
\label{VExpansion}
\end{equation}
with  constant coefficients $b_{ij,p}$. We therefore see that  in the vicinity of the pole   the scalar potential decomposes into an exponential fall-off from a de Sitter plateau as
\begin{equation}\label{eq:LOplateauV}
V=V_0\left(\theta_i,\psi_\beta\right)-e^{-\sqrt{\frac{2}{3\alpha}}\varphi}\, V_1(\theta_i,\psi_\beta)+{\cal O}\left(e^{-2\sqrt{\frac{2}{3\alpha}}\varphi}\right)\, ,
\end{equation}
with $V_0$ and $V_1$ functions of the angular variables as dictated by eq.~\eqref{VExpansion}. It is interesting to note that the residue of the pole does not depend on either $A_i$ or $\Omega_i$, leading the exponential plateau to have a universal nature. The slope of the exponential fall-off is therefore not affected by the particular radial direction in field space. The amplitude of the plateau is exclusively determined by the angular and axionic  directions $(\psi_\beta,\theta_i)$. This effect can be observed in fig.~\ref{plots}.

If inflation occurs purely in the $R$ direction, the observable predictions of this model, regardless of the inclusion of multiple fields, retain the universality properties extensively discussed in the literature (see e.g. \citep{Kallosh:2013yoa,Galante:2014ifa,Broy:2015qna,Scalisi:2015qga}) for the single field case, 
\begin{equation}
n_s=1-\frac{2}{N_e}\, ,\qquad r=\frac{12 \alpha}{N_e^2}\, ,
\label{predictions}
\end{equation}
where $N_e$ is the number of e-folds of expansion between the pivot scale leaving the horizon and the end of inflation. 
However, the angular and axionic fields might play a role in the inflationary dynamics leading to multifield effects that can modify the predictions. To assess this, we need to study the hierarchies in the mass spectrum.

\subsection{Scaling of mass spectrum}\label{sec:mass-spectrum}

The hierarchies in the mass spectrum are intimately related to the eigenvalues of the field space metric given by eq.~\eqref{KMang1} and  eq.~\eqref{eq:kinetic}. Using the canonically normalised field $\varphi$ we define the parameter $\epsilon$ as a measure of proximity to the moduli boundary:
\beq
R=\tanh(\varphi/\sqrt{6\alpha})\simeq 1 - e^{-\sqrt{\frac{2}{3\alpha}}\varphi}\equiv 1 - \epsilon\,.
\label{canR}
\eeq
We can then see that the line element scales with the proximity to the boundary as
\begin{equation}
  -\dfrac{1}{2}(\partial \varphi)^2-\frac{3\alpha}{2\epsilon} \sum_{i}(\partial_{\beta}\Omega_i)^2 \, \partial \psi_\beta\partial \psi^\beta - 3\alpha \sum_{ij} \left[\frac{1}{4\epsilon^2}\Omega^2_i \Omega^2_j+ \frac{1}{2\epsilon} \delta_{ij}\Omega^2_i \right] \partial \theta_i \partial \theta_j \,.
\end{equation}
where we have neglected higher order contributions in $\epsilon$. 

In general, it is not straightforward to compute the eigenvalues of this kinetic metric but we can consider specific field configurations which simplify the situation. We are particularly interested in the regime where all the fields $\Phi_i$ contribute equally when approaching  the boundary, \emph{i.e.} when all the branes are equally displaced from the origin and the inflationary dynamics is determined by their collective motion. In this maximally multifield case, this corresponds to the choice   $\Omega_i^2=1/N$ for any $i$. In this case, following Appendix~\ref{angles}, the metric element for the angular coordinate $\psi_\beta$ is
\beq
-\frac{3\alpha}{2\epsilon}\sum_{i} (\partial_{\beta}\Omega_i)^2 = -\frac{3\alpha(N-\beta + 1)}{2 N \epsilon},
\label{psi_ev}
\eeq
which ranges from $3\alpha/2\epsilon$ to $3\alpha/N\epsilon$. 

\begin{figure}
\centering
 \includegraphics[width=50mm]{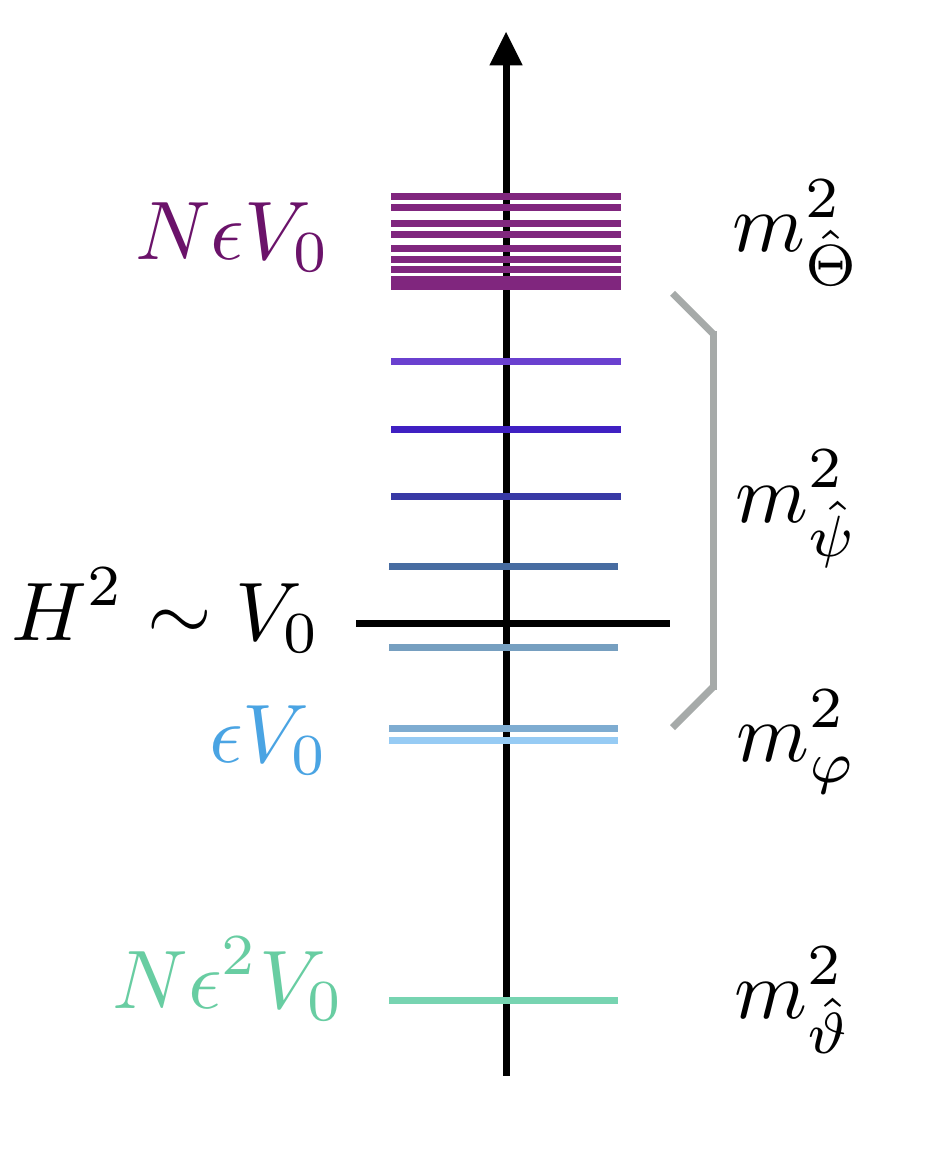}
\caption{Mass hierarchies of eq.~\eqref{mspect} in the deep plateau limit, $\varphi\gg 1$, and large $N$ limit.}
\label{mass_spectrum}
\end{figure}

Also in this regime, close to the boundary, the metric for the axionic fields $\theta_i$ eq.~\eqref{eq:kinetic} can be written as 
\beq\label{eq:fa}
 \ G_{ij} =     v^2 J_{ij}+\delta_{ij}v \ \quad , \quad v= \dfrac{R^2}{N(1-R^2)} \simeq  \dfrac{1}{ 2 N \epsilon} \ ,
\eeq
where $J_{ij}$ is the all-ones matrix (a square matrix with   all entries equal to 1). The eigenvalues of this metric are easy to compute: the matrix $J_{ij}$ has rank one, with one single non-zero eigenvalue equal to  $\text{ Tr} (J_{ij} )= N$. The identity matrix is invariant under any transformation and therefore $G_{ij}$ has all but one eigenvalue equal to $v$. We denote the corresponding eigenvectors of the $N-1$ equal eigenvalues, $\Theta_a$, with $a=1, \, \ldots \, , N-1$. The last eigenvalue, corresponding to what we define as the $\vartheta$ direction, is
\beq\label{eq:ftilde}
Nv^2+v=\dfrac{1}{N}\dfrac{R^2}{(1-R^2)^2} \simeq \frac{1}{4 N \epsilon^2} \ .
\eeq
At the point in field space where the metric has these eigenvalues, we can locally canonically normalize the angular fields by defining
\begin{equation}
\begin{aligned}
&\hat\psi_\beta \equiv \sqrt{\frac{3\alpha(N-\beta + 1)}{ N \epsilon}}\psi_\beta\ \\
&\hat\Theta_a \equiv \sqrt{\frac{3\alpha}{N\epsilon}}\Theta_a  \\
&\ \hat{\vartheta} \ \equiv \sqrt{\frac{3\alpha}{2N\epsilon^2}}\vartheta \, 
\end{aligned}
\end{equation}
and write the potential near the pole as
\begin{equation}
V=V_0\left(\hat\Theta_a,\hat\vartheta,\hat\psi_\beta \right)-\epsilon \, V_1\left(\hat\Theta_a,\hat\vartheta,\hat\psi_\beta \right)+{\cal O}\left(\epsilon^2\right)\, .
\end{equation}
Owing to the local canonical normalization of the kinetic terms, the mass spectrum therefore scales as
\begin{equation}
\begin{aligned}
&\left| V_{\varphi\varphi} \right| \sim \epsilon V_{1} \sim \epsilon V_0  \\
&\left| V_{\hat\psi_\beta\hat\psi_\beta}  \right| \sim  \frac{N \epsilon}{N-\beta + 1} V_0  \\
&\left| V_{\hat\Theta_a\hat\Theta_a}  \right| \sim N \epsilon V_0 \\
&\left| V_{\hat\vartheta\hat\vartheta}  \right| \sim N \epsilon^2 V_0  
\label{mspect}
\end{aligned}
\end{equation}
for a generic scalar potential where usually ${\mathcal{O}}(V_0) \approx {\mathcal{O}}(V_1)$. The mass scaling of the $N-1$ elliptical angular fields $\hat\psi_{\beta}$ ranges from $\epsilon V_0$ to $N \epsilon V_0$, \emph{i.e.} from the same scaling as the light radial field $\varphi$ to the scaling of the heavier axionic fields $\hat\Theta_a$.
We should therefore allow for the possibility that some of these fields might contribute to the inflationary dynamics. These effects might lead to multifield deviations from the simplest predictions given by eq.~\eqref{predictions}. The $N-1$ axions $\hat\Theta_a$, in the large $N$ limit, correspond to a heavier sector which we do not expect to contribute significantly to the dynamics. The single $\hat\vartheta$ direction, in the limit of  $\varphi\gg 1$, is exponentially lighter than all other sectors and becomes a true spectator; in this deep plateau limit the dynamics of $\hat\vartheta$ is frozen in deep slow-roll and will resemble the case of the single angular field of ref.~\cite{Achucarro:2017ing}. The kinetic scaling of the mass spectrum of the theory for a configuration where all fields contribute equally to inflation is illustrated in Figure~\ref{mass_spectrum}.

\section{The fate of inflationary plateaus - weak gravity strikes back}\label{Sec:WGC}

Effective field theories arising from a theory of quantum gravity  are constrained by consistency conditions such as the  Weak Gravity Conjecture (WGC)~\cite{ArkaniHamed:2006dz} or the Swampland Distance Conjecture (SDC)~\cite{Ooguri:2006in,Ooguri:2016pdq,Freivogel:2016qwc,Brennan:2017rbf}. The SDC states that whenever one moves  an infinite  distance  in moduli space, an infinite tower of states   becomes massless causing the break of the effective description. We will comment on the connection between the SDC and $\alpha$-attractor models later in this section, but for now we focus on enforcing consistency arguments coming from the weak gravity conjecture.

The WGC arose as a proposal to argue that black holes should always be able to evaporate via Hawking radiation, such that the final state of any charged black hole would be able to decay and leave no remnant. For this to happen it is necessary that any theory of quantum gravity has at least a fundamental object fulfilling  the condition 
\beq
1\lesssim \dfrac{M_P}{m}q  \label{eq:morelables}
\eeq
 such that the decay process is possible for any black hole. A number of attempts were made to further constrain the particular object fulfilling the WGC condition. These have given rise to several versions of the WGC. In particular, the strong form of the WGC requires the lightest particle on the spectrum to be the one fulfilling the above condition, whereas more loose forms do not impose further conditions on the particular particle fulfilling the condition.  It is still unknown which version of the conjecture (if any) is the right one and it is not our intention to provide any new insight in this direction, so we just refer the interested reader to~\cite{Rudelius:2014wla,Rudelius:2015xta,Montero:2015ofa,Brown:2015iha,Hebecker:2015rya,Brown:2015lia,Bachlechner:2015qja,Heidenreich:2015wga,Heidenreich:2015nta,Ibanez:2015fcv,Hebecker:2015zss,Baume:2016psm,Montero:2016tif,Cottrell:2016bty,Heidenreich:2016aqi,Klaewer:2016kiy,Hebecker:2016dsw,Hebecker:2017wsu,Hebecker:2017uix,Landete:2017amp,Palti:2017elp,Ibanez:2017oqr,Hamada:2017yji,Montero:2017mdq,Hebecker:2017lxm,Valenzuela:2017bvg,Ibanez:2017vfl,Lust:2017wrl,Heidenreich:2017sim,Andriolo:2018lvp} for extensive discussion. Our purpose, instead, will be to apply the WGC constraints to the $\alpha$-attractor models as a consistency requirement to allow for a string theory embedding of these effective field theories.  For doing so  we will consider the sublattice WGC \cite{Heidenreich:2016aqi} (see also the related tower WGC \cite{Andriolo:2018lvp}) which is supported by evidence from string theory. According to this version of the WGC,   a sublattice of the entire charge lattice allows for the decay of an extremal black hole to happen.

Since string theory contains fundamental charged objects of different dimension, first note that  the above  argument can be extended to other black objects of different dimensionality. This implies one is not completely free concerning the assignment of charges and tensions of the fundamental objects of any theory.  For our purposes we will be interested in how the WGC applies to instantons. In this case, the WGC sets the bound
\beq \label{eq:wgc-inst}
\mathcal{S}_E\lesssim \dfrac{\Mp}{f}n \ ,
\eeq
where $\mathcal{S}_E$ is the euclidean action of the instanton, $f$ the decay constant of the axion coupled to it and $n$ the instanton number/charge. This bound will be our starting point to set the limitations of $\alpha$-attractors.  To do so while working in a controlled effective field theory, we want to guarantee a controlled instanton expansion. This typically requires  $\mathcal{S}_E\simeq n$ for the charge-$n$ instanton.   Recalling that we are taking $\Mp=1$, the above WGC bound, when applied to the single complex field case (single axion) reads
\beq
f\lesssim 1 \quad .  \label{eq:aajhsdfvu}
\eeq
This is in agreement with evidence from certain classes of string theory compactifications~\cite{Banks:2003sx}.


 The above inequality   follows from demanding that every instanton with charge $n$ 
satisfies $\mathcal{S}_E\simeq n$. 
  However, while control of the instanton expansion clearly forbids violating $\mathcal{S}_E\simeq n$ for a large number of instantons, conceivably control might remain feasible if e.g. a single charge-$n$ instanton satisfies a weaker bound $\mathcal{S}_E\simeq \tilde n,  $  where $ \tilde n< n$.     This leads to  a milder  constraint   than (\ref{eq:aajhsdfvu}) since combined with the WGC it leads to the bound 
$ f\lesssim  \frac{n}{\tilde n} \ (\equiv n_{eff} > 1$, note that $n_{eff}$ is rational and not neccessarily natural) for the instanton dominating the axionic potential.  Note that this is possible only if the instanton relevant for the WGC has $n>1$, a possibility allowed by the sublattice WGC \cite{Heidenreich:2016aqi}.  This loosening essentially resembles the 
 `loophole' in~\cite{Rudelius:2015xta,Brown:2015iha}.   

For effective field theories including multiple axions, the above argument needs to be extended. In the black hole picture this corresponds to a theory with multiple $U(1)$ gauge factors, first studied in ref.~\cite{Cheung:2014vva}. The key factor in this case is that black holes can be charged under more than a single $U(1)$ factor at the same time, and thus the WGC constraints cannot be implemented by considering each of these $U(1)$ factors separately. One needs to consider the lattice of charge-to-mass ratios of the theory such that the single $U(1)$ bound extends to the so-called \emph{convex hull condition} on this lattice. This condition was first   translated to the axion language in ref.~\cite{Rudelius:2014wla}.  Before discussing this condition in more detail, we note that in what follows we will consider all instantons to be relevant in the sense discussed above.

The convex-hull condition is nothing but a basis independent (in axion space) generalization of  eq.~(\ref{eq:wgc-inst}). Let us consider first the typical criterion for control over the instanton expansion that lead us to (\ref{eq:aajhsdfvu}).  In order to achieve basis independence, we proceed by considering a $N$-dimensional polygon whose edges are located at positions
\beq
\pm (0, ... ,0 ,  1/f_i , 0,  ..., 0) \quad , \quad i=1, ... , N \ ,
\eeq
 $f_i$ being the decay constants of the  $N$ axions in the model under consideration.  The condition eq.~(\ref{eq:aajhsdfvu}) applied to each of these vectors requires all edges to sit at least at distance 1 from the origin. This is a basis dependent statement, so demanding instead that the whole polygon contains the unit ball gives rise to basis independence. This geometric requirement is quite simple but not very practical. Luckily, it can be translated into a more practical and intuitive inequality  \cite{Rudelius:2014wla}:
\beq
\sum_if^2_i\lesssim 1 \ . \label{eq:easy2}
\eeq
Again, we find a constraint on the possible decay constants of axions, but this time the constraint involves  all axions relevant for the WGC at the same time.   We can now easily soften this constraint as done for the single axion case by allowing for some instanton(s) with smaller instanton action such that the mild version of the above inequality reads 
\beq
\sum_if^2_i\lesssim n_{eff}^2 \ . \label{eq:easy3}
\eeq

Since we are setting bounds on axion decay constants, we next argue that the fields $\theta_i$ in \S\ref{Sec:Uni} are indeed axions. It is nowadays standard in supergravity to denominate axions those fields that do not appear  in the K\"{a}hler potential and whose potential is periodic.  In order to apply WGC arguments, these conditions are necessary but not sufficient: one needs to argue that the axion potential can only be generated by instantons. In order to do so, recall that the string theory picture corresponding to our inflationary setup corresponds to the motion of D3-branes, whose position we parametrized by the fields $\Phi_i$. It is known that in standard type IIB compactifications \textit{\`{a} la} GKP \cite{Giddings:2001yu,Grimm:2004uq}, the radial part $|\Phi_i|$ of the position moduli is massless at the perturbative level unless supersymmetry is broken, while their axion phases $\theta_i$ remain flat directions always in perturbation theory. So,  it is necessary to also include non-perturbative objects on the compactification in order to stabilize the $\theta_i$. The inclusion of Euclidean D3-branes will indeed generate a potential and stabilize these moduli. In the 4D supergravity language these instantons generate a non-perturbative superpotential \cite{Witten:1996bn,Baumann:2006th}
\beq
W_{np}= F(\Phi_i, z_I )e^{- T_J}  ,
\eeq 
$T_J$ being the superfield describing   the size of $\Sigma_J$ where the ED3-brane is wrapped and $F $ a holomorphic function of the D3-brane moduli as well as  other geometric moduli $z_I$, such as the complex structure ones.  Note here that holomorphy of the superpotential will ensure  the potential is compatible with the axion periodicity arising from our choice of polar coordinates $\theta_i \sim \theta_i+2\pi$. Using this fact, we conclude that the potential of the $\theta_i$ is  indeed generated by instantons and compatible with the usual identification coming from each axion living on a $\mathbf{S}^1$.\footnote{We note that the 4D low-energy effective description of instantons from string theory reduces in many cases to the Giddings-Strominger type gravitational instantons~\cite{Giddings:1987cg,ArkaniHamed:2007js}. Expanding the axion on a $n$-instanton background provides the usual axion-instanton term enforcing   $\theta_i\sim\theta_i+2\pi$ for the path integral to be well defined. }

 As a final step before  applying WGC constraints, note that it is always possible to rescale the axions $\hat{\theta}_i= f_i\theta_i$, such that  if the periodicity of $\theta_i $ is given by $\theta_i\sim\theta_i+2\pi$, then for $\hat{\theta}_i$ it is $\hat{\theta}_i\sim\hat{\theta}_i+2\pi f_i$. Therefore the decay constant changes with rescaling, and so does the axion kinetic term. In the single field case, the   relevant scale for WGC arguments is the one giving rise to a canonical kinetic term for the axion, or equivalently, the square root of the prefactor in the kinetic term for an axion with periodicity $\theta_i\sim\theta_i+2\pi$. In the multiple axion case, we need to apply the same criterion to each axion in a basis where there is no kinetic mixing. Due to our initial  coordinate choice where $\theta_i\sim\theta_i+2\pi$ for all axions, this task can easily be carried out: we just need to compute the eigenvalues of the field space metric. In fact,  after a change of coordinates the  kinetic Lagrangian will be $-\sum_i \frac{f_i^2}{2}\partial_\mu\theta_i\partial^\mu\theta_i$ which is canonically normalized by the change $\hat{\theta}_i= f_i\theta_i$, such that $f_i$  are the decay constants of interest for WGC arguments.

We are now ready to start  studying the consequences for the single (complex) field case.  In this case, the K\"ahler metric is
\beq
 K_{\Phi \bar \Phi}=\frac{3\alpha}{ (1- \Phi\bar\Phi)^2}\,,
\eeq
where $ \Phi\bar\Phi = \phi^2$, and leads to an axionic partner of $\phi$ with a decay constant
\beq
f^2 (\phi )= 6\alpha\dfrac{\phi^2}{(1-\phi^2)^2}   \ .
\label{single_f}
\eeq
As the field $\phi$ approaches the moduli boundary, the decay constant of the axionic partner diverges; this behaviour has been noticed in ref.~\cite{Linde:2018hmx}. The WGC therefore constraints the maximum displacement in the radial direction. To understand how stringent this constraint is, it is interesting to compare the WGC bound $f^2\lesssim n_{eff}^2 \ $ with the slow-roll condition $\epsilon_{\rm SR} <1$. Assuming inflation occurs primarily in the radial direction and following eq.~\eqref{KMang1}, we have that in the slow-roll limit
\beq
\epsilon_{\rm SR}  \simeq \frac{1}{h^2} \left(\frac{V_{\phi}}{V}\right)^2  \, , \quad h^2=3\alpha\frac{1}{(1-\phi^2)^2} \, .
\eeq
Taking the potential to be expanded as in eq.~\eqref{VExpansion} and dominated by ${\mathcal O}(\phi^2)$ terms, inflation occurs when
\beq
 3\alpha\dfrac{\phi^2}{(1-\phi^2)^2} > 1   \, ,
\eeq
which comparing with eq.~(\ref{single_f}) is equivalent to $f^2 \gtrsim {\cal O}(1)$.  This implies that obtaining sufficient inflation in this scenario, while remaining compatible with the WGC, requires $n_{eff}$ to be quite large, which is rather unnatural. 
If instead one considers the strong bound from the WGC where $n_{eff}=1$, then $\alpha$-attractors turn out to be unable to provide enough inflation.  This observation is independent of the value of $\alpha$.  From here we conclude that the simplest supergravity $\alpha$-attractor model, when embedded and coupled to quantum gravity, is  in direct conflict with consistency requirements coming from the   weak gravity conjecture, which in turn is supported by evidence from string theory \cite{Banks:2003sx}.

We now  apply a similar argument to the $N$ field case studied above. We saw in \S\ref{Sec:Uni} that, taking polar coordinates for the complex field, the field-space metric on the axionic sector is not diagonal. In order to apply the WGC bound we choose a configuration where the inflationary dynamics is carried out by the collective motion of all D3-branes, \emph{i.e.} when all fields contribute equally and all angular functions $\Omega_i$ have the same value $\Omega_i^2=1/N$. 
Diagonalizing the field space metric in this configuration gives rise to the axionic  fields $\Theta_a\,,\,a=1\ldots N-1$, and  $\vartheta$ with diagonal  kinetic terms 
\beq\label{eq:ThetaAction}
  - \sum_a \frac{f_a^2}{2}\partial_\mu\Theta_a\partial^\mu\Theta_a - \frac{ f_\vartheta^2}{2}\partial_\mu\vartheta\partial^\mu\vartheta\ .
\eeq
The  decay constants are given by eqs.~(\ref{eq:fa}) and (\ref{eq:ftilde})\footnote{We note here that, in $\alpha$-attractor models, inflation does not involve an active axion or linear combination of axions. Therefore, the only input needed from the axionic sector of the theory are the periodicities of the canonically normalized axions. This is unlike models where inflation occurs in the axionic sector, such as $N$-flation \cite{Dimopoulos:2005ac}, where it is necessary to have information about the instanton numbers on each cycle in order to compute the potential and 
the fundamental field-space domain of the axions; see e.g.  \cite{Bachlechner:2014gfa,Bachlechner:2015qja} for a discussion of  models of this type. } 
\beq\label{eq:decayconstants}
f_a^2  = \dfrac{6\alpha R^2 }{N(1-R^2)} \quad  \,  ,   \quad f_{\vartheta}^2 = \dfrac{6\alpha R^2 }{N(1-R^2)^2} \;\ .
\eeq 
To study the bound the WGC sets on this model, we plug these expressions in eq.~(\ref{eq:easy3}) to find:
\beq
f_{\vartheta}^2 + (N-1) f_{a}^2=\dfrac{1}{N}\left( \dfrac{6\alpha  R^2 }{(1-R^2)^2} - \dfrac{6\alpha  R^2 }{(1-R^2)}\right)+\dfrac{6\alpha  R^2 }{(1-R^2)}  \lesssim n^2_{eff} \ .
\eeq
When $N=1$ this expression reduces to the result derived above, whereas for large $N$ only the last term is relevant. Na\"ively we might have expected that the inclusion of the assistance of $N$ fields could relax the WGC bound allowing the system to be sufficiently displaced in $R$ while keeping away from the moduli boundary. However, we can see by the $N$-independence of the last term that this does not happen. In fact, it turns out that the large $N$ limit leads to a slightly weaker bound on the maximum allowed value for $R$, but the gain turns out to be negligibly small. This puts pole N-flation in a situation similar to the single field case: strong bounds from  the WGC are incompatible with the production of sufficient inflation. Only weaker bounds where the relevant instantons lead to   unnaturally large    $n_{eff}$ would be able to provide enough e-folds to render these models viable.

We would like to contrast this with the case of axion N-flation. As previously discussed in the literature (see e.g. \cite{Rudelius:2015xta}), diagonal enhancements in axion N-flation are incompatible with the   WGC: the basis independence described by the convex hull condition results in the cancellation of the proposed $\sqrt{N}$ enhancement in this  direction in axion space, such that the maximum allowed displacement is independent of direction.
As we argued above, we also found little gain in this regard when applying the   WGC to pole N-flation. 

But if some higher instanton(s)   are the relevant
ones for the WGC as described above, there exists the possibility of (slightly) larger axion decay constants. This results in a little gain for the available inflaton displacement in the case of axion N-flation, but for pole N-flation it relaxes the bound on the radial field displacement such that the inflaton can (again slightly) reach the plateau area.


 A difference between pole N-flation and axion N-flation arises from the fact that in pole N-flation the WGC constraints are related to  $f$'s but the particular shape of the axion potential coming from the instantons involved is not relevant while for axion N-flation these instantons play a crucial role in shaping the inflaton potential.

As a final remark, we would like to make a connection between these results and the swampland distance conjecture (SDC). Consider a very loose version of the bound imposed by the WGC, with $n_{eff}\gg 1$, thus  allowing the inflaton to get deep into the plateau. In this case when the inflaton approaches the boundary (when $R\rightarrow 1 $) not only do the decay constants grow exponentially according to eq.~(\ref{eq:decayconstants}), but also the masses of the axions become exponentially small as can be seen in eq.~(\ref{mspect}). As the plateau itself arises from a kinetic term with a 2nd order pole, this  behaviour shows certain similarity to recent arguments in favor of the SDC provided in ref.~\cite{Grimm:2018ohb,Heidenreich:2018kpg} as reaching the 2nd order pole of the metric on moduli space makes the axions in our setup exponentially light.

 As a consequence of these observations, we conclude that the WGC rules out the \textit{infinite} inflationary plateaus of pure supergravity $\alpha$-attractor models based on disk variable-type kinetic terms with 2nd order poles. This suggests that Pole N-flation with infinite plateaus does not admit a UV completion in string theory.

\section{Towards pole N-flation in type IIB string theory}\label{Sec:FibCY}

As we have seen in the previous sections, the K\"ahler potential \eqref{Kvm} becomes singular exactly at the same point where the kinetic Lagrangian develops a pole. This fact poses strict limitations on the possibility of realizing the pole inflation scenario within a supergravity framework. Indeed, the F-term scalar potential ${V=e^K (\ldots)}$  will in general not be regular at this point in field space and the inflationary plateau will be easily spoiled. To avoid this situation, one can tune the superpotential such as to cancel the pole induced by the exponential pre-factor in $V$, but this appears to be a quite non-generic and model-dependent situation. But even granted this possibility, we will interpret the field approaching the pole as a shrinking volume of extra dimensions in string theory. If this volume is the total volume of the extra dimensions, sending this to zero will send perturbative corrections soaring in magnitude and thus compromising control.

A rather more appealing alternative is to find a class of models where the form of $K$ has a regular behaviour while still inducing a pole in the corresponding kinetic structure. Interestingly, stabilizing the overall volume of fibred Calabi-Yau (CY) geometries~\cite{Cicoli:2008gp,Broy:2015zba,Cicoli:2018tcq} using the Large Volume Scenario (LVS) mechanism \cite{Balasubramanian:2005zx} provides a large class of string models with a K\"ahler potential with the desired properties. 

In the following, we will review the main characteristics of this framework and show how to embed the pole N-flation picture therein.  We will also discuss moduli stabilization of this setup, pointing out its limitations given the current status of knowledge on quantum corrections. Finally, we will provide an analysis of the model's dynamics and cosmological predictions.

\subsection{Pole N-flation from fibred Calabi-Yau manifolds}

A large fraction of CY manifolds are $K3$-fibred. This means that the positive part of the CY volume takes the form
\beq
\mathcal{V}=\kappa_{122} v^1 (v^2)^2\,,
\eeq
in terms of the 2-cycle volumes $v^i$, and $\kappa_{122}$ is the intersection number between the 2-cycles on the given Calabi-Yau manifold. The 4-cycle volumes $\tau_i$ are related to the 2-cycles by
\beq
\tau_i =\frac{\partial\mathcal{V}}{\partial v^i}\,,
\eeq
allowing us to write the volume of a $K3$-fibred CY in terms of the 4-cycle volumes as
\beq
\mathcal{V}\sim \sqrt{\tau_1}\tau_2\,.
\eeq
The corresponding K\"ahler potential then takes the form
\beq
K=-2\ln\,\mathcal{V}=-\ln(T_1+\bar T_1)-2\ln(T_2+\bar T_2)\,.
\label{original}
\eeq
where we have introduced the volume moduli $T_i$, which are related to the 4-cycle volumes by means of $2 \tau_j = T_j + \bar{T}_j$ while their axionic partners are $2c_j=(T_j-\bar T_j)/i$. 

Now assume the CY to possess a warped near-conifold region. Assume further that the 4-cycle $\Sigma^4_2$ with volume $\tau_2$ reaches somewhat into the warped region. This is not particularly restrictive, as we can stabilize  part of the complex structure moduli using flux near conifold points for a large fraction of all $K3$-fibred CYs. Finally, place a number $N$ of $D3$-branes at the IR end of the warped region.

The K\"ahler potential for models in this class will look like
\beq
K=-\ln \left( T_{1}+\bar{T}_{1}\right) -2\ln \left( T_{2}+\bar {T}_{2}-R^2\right)\,,
\label{fibered_Npole}
\eeq
where we define as before
\beq
R^2\equiv\sum ^{N}_{i=1}A_{i}\Phi _{i}\bar {\Phi }_{i}\,,
\eeq
with $\Phi_i$ parametrizing the positions of the $D3$-branes. The O7-orientifolding enforces the relation between 2-cycle volumes, 4-cycle volumes and D3-brane coordinates such that~\cite{Grimm:2004uq,Martucci:2014ska} 
\beq
\tau_1=(v^2)^2   \quad , \quad   \tau_2=v^1v^2 + \frac{R^2}{2}  \quad \Rightarrow \quad 
v^2=\sqrt{\tau_1}  \quad , \quad v^1=\frac{1}{\sqrt{\tau_1}}(\tau_2-\frac{R^2}{2})\ .
\label{cycle_relation}
\eeq
The corresponding expression for the CY volume now reads
\beq
\vol \sim \sqrt{\tau_1}(\tau_2-\frac12 R^2)\,.
\eeq
An alternative construction might instead shift $\tau_1$ by the D3-brane K\"ahler potential $R^2/2$. In this case, the CY volume would become 
$\vol \sim \tau_2\sqrt{\tau_1-R^2/2}$.

The LVS scheme of volume stabilization can now proceed if we assume the total CY to have a third pure blow-up K\"ahler modulus $\tau_3$, such that the CY volume becomes
\beq
\vol \sim \sqrt{\tau_1}(\tau_2-\frac12 R^2)- \lambda_3\tau_3^{3/2}\,.
\eeq
We therefore include the leading order type IIB $\alpha'$-correction into the K\"ahler potential
\beq
K=-2\ln(\vol+\xi/2)\,,
\eeq
and $\tau_3$ acquires an ED3 instanton contribution in the superpotential, in addition to the constant piece from 3-form fluxes, such that
\beq
W=W_0+A e^{-2\pi T_3}\,.
\eeq
This setup will stabilize the modulus $\tau_3$ and the whole leading-order volume combination ${\vol_0\equiv\sqrt{\tau_1}(\tau_2-\frac12 R^2)}$ at VEVs with a relation
\beq
\langle\vol_0\rangle\sim e^{2\pi\langle\tau_3\rangle}\,.
\eeq

In order to reproduce the pole N-flation dynamics, schematically encoded by eq.~\eqref{Kpot}, we would like to stabilize the modulus $\tau_2$ \emph{separately}. For this purpose, we first observe that the scales of LVS stabilization operate at ${\cal O}(\vol^{-3})$. This rules out the possibility of stabilizing $\tau_2$ supersymmetrically \emph{\`a la} KKLT, by adding a non-perturbative effect to $W$. The resulting potential terms from the KKLT mechanism would indeed appear at ${\cal O}(\vol^{-2})$ and eventually spoil the LVS mechanism.

Hence, we need to stabilize $\tau_2$ perturbatively, presumably using an interplay of string loop corrections and higher-order $F$-term contributions to the scalar potential, which operate starting at ${\cal O}(\vol^{-10/3})$. However, in the known simple cases, where we can compute some of the string loop corrections to $K$ and the $F^4$-terms in the scalar potential~\cite{Ciupke:2015msa}, these depend on the 2-cycle volumes $v^i$~\cite{vonGersdorff:2005bf,Berg:2005ja,Berg:2007wt,Cicoli:2007xp}. Therefore, looking at expressions \eqref{cycle_relation}, these corrections do not affect $\tau_2$ individually but rather $\tau_1$ and the whole combination $\tau_2-R^2/2$.

At this point, we  content ourselves with merely pointing out as a challenge the need to explicate a perturbative stabilization mechanism which will stabilize $\tau_2$ just by itself. From now on, we will simply assume that such stabilization for $\tau_2$ exists. 

As a final remark, we wish to point out that we could have instead looked at the case where the whole combination $\tilde\tau_2\equiv\tau_2-R^2/2$ is given a potential and is stabilized by string loop corrections such as those discussed above. For those models, one can show that the structure of the kinetic terms, in terms of $\tau_1$, $\tilde\tau_2$, $R$ and the angular variables $\psi_\alpha,\theta_i$, reduces to ${\cal L}_{kin.}=-\frac{1}{4\tau_1^2}(\partial\tau_1)^2-\frac{1}{2\tau_2^2}(\partial\tilde\tau_2)^2- \frac{1}{\tilde\tau_2}(\partial R)^2 +{\cal L}_{kin.}(\partial\psi_\alpha,\partial\theta_i)$. If the potential only has contributions of the type discussed above, this setup resembles precisely the original fibre inflation setup~\cite{Cicoli:2008gp} (see also~\cite{Broy:2015zba,Cicoli:2016chb}) in terms of the effective half-plane variables $\tau_1,\tilde\tau_2$~\cite{Kallosh:2017wku} -- except for the extra $2N$ massless spectator fields: $2N-1$ angular fields $\psi_\alpha$ and $\theta_i$ and one field direction given by a linear combination of $R$ and $\tau_2$ orthogonal to $\tilde\tau_2$. If in general these $2N$ fields are also given a potential, we expect a rich mass spectrum and possible multifield phenomenology in analogy with \S\ref{Sec:Uni}. In this work we do not study this type of model, focusing instead on the stabilized   $\tau_2$ case.

\subsection{Dynamics of fibred pole N-flation and universal predictions}

The effective K\"ahler potential of fibred pole N-flation reads
\beq
K=-\ln \left( T_1+\bar T_1\right) -2\,\ln \left(2\langle\tau_2\rangle\right) -2\,\ln \left( 1-\frac{R^2(\Phi_i,\bar{\Phi}_j)}{2\langle\tau_2\rangle}\right)\,,
\eeq
once we assume stabilization of $\tau_2$. Note that the last contribution is identical to eq.~\eqref{Kpot}, with $3\alpha =2$ and up to a multiplicative factor in $R$. Therefore in the following analysis we can employ the results derived in \S\ref{Sec:Uni}.

The kinetic Lagrangian of the dynamical degrees of freedom is given by 
\beq
-\left.K_{T_2\bar T_2}\right|_{\tau_2=\langle\tau_2\rangle}(\partial c_2)^2-K_{T_{1}\bar {T}_{1}}\partial T_{1}\partial \bar {T}_{1}-K_{\Phi _{i}\bar {\Phi }_{i}}\partial \Phi _{i}\partial\bar{\Phi }_{j}\,.
\eeq
Applying the LVS procedure for volume stabilization forces ${2\tau_1=T_1+\bar T_1}$ to be a function of $R$, such as 
\beq
\tau _{1}(R)=\frac {\mathcal{V}_0^2}{\langle\tau _{2}\rangle^2}\,\frac{1}{\left( 1-\frac{R^2}{2\langle\tau_2\rangle}\right)^2}\,,
\eeq
with $\mathcal{V}_0$ being the stabilized volume.  This implies an additional contribution to the total kinetic term of $R$ of the form
\bea\label{kinRtau1}
-\frac{1}{(2\tau_1)^2} \left(\frac{\partial\tau_1}{\partial R}\right)^2(\partial R)^2 = - \frac{R^2}{\langle\tau _{2}\rangle^2} \frac{1}{\left(1-\frac{R^2}{2\langle\tau _{2}\rangle}\right)^2}(\partial R)^2\,.
\eea
Therefore, after volume stabilization, the field-space metric for the radial direction is determined by eq.~\eqref{KMang1} together with the contribution of the D3-branes from eq.~\eqref{kinRtau1} (with the $R$ properly rescaled):
\bea
-\left[\frac{R^2}{\langle\tau _{2}\rangle^2} \frac{1}{\left(1-\frac{R^2}{2\langle\tau _{2}\rangle}\right)^2} + \frac{1}{\langle\tau _{2}\rangle\left(1-\frac{R^2}{2\langle\tau _{2}\rangle}\right)^2} \right] (\partial R)^2=- \frac{R^2+\langle\tau _{2}\rangle}{\langle\tau _{2}\rangle^2\left(1-\frac{R^2}{2\langle\tau _{2}\rangle}\right)^2}(\partial R)^2\,. \label{totalkinR}
\eea

In order to absorb the $\langle\tau _{2}\rangle$ dependence, we define $\tilde R \equiv R/ \sqrt{2\langle\tau _{2}\rangle}$ such that the kinetic term becomes
\begin{equation}
-\frac{2 (1 + 2\tilde R^2)}{(1 - \tilde R^2)^2}\  (\partial \tilde R)^2\,.
\end{equation}
This allows us to define the canonically normalized field $\varphi$ corresponding to the radial field $R$ as
\beq
d\varphi \equiv 2\frac{\sqrt{1+2\tilde R^2}}{1-\tilde R^2}d\tilde R\quad.
\eeq
We see that $\tilde R \to 1$ corresponds to $\varphi\to\infty$, which as done in \S\ref{sec:mass-spectrum} we can use to express $\varphi$ in terms of $\epsilon=1-\tilde R$
\beq
d\varphi = -\frac{\sqrt 3}{1- \tilde R}+{\cal O}(1)
\eeq
such that 
\beq
1-\tilde R = \epsilon =  e^{-\frac{\varphi}{\sqrt 3}}\quad.
\eeq
This expression is precisely eq.~\eqref{canR} for $\alpha=2$. Using analogous arguments to those in \S\ref{Sec:Uni}, we make an expansion of the scalar potential as eq.~\eqref{VExpansion}.
This generic structure of the scalar potential of $\Phi_i$ as a power law series around its minimum often arises for open string moduli in setups with controlled moduli stabilization and supersymmetry breaking. For example, refs.~\cite{Baumann:2007ah,Baumann:2010sx} argue explicitly that mobile D3-branes at the IR end of the warped throat of a KKLT or LVS compactification acquire a scalar potential of the general form of eq.~\eqref{VExpansion}. This results into a computable spectrum of discrete values of $p\geq 1$ while the coefficients $a_{i,p}\;,\; b_{ij,p}$ are tunable Wilson coefficients except the one arising from the conformal curvature coupling of the D3-brane moduli. 

Finally, in analogy with \S\ref{Sec:Uni}, if the motion is purely radial, we see that for an arbitrary number $N$ of open string moduli $\Phi_i$ driving exponential plateau inflation, we arrive at 
\beq
\alpha = 2\quad\Rightarrow\quad n_s=1-\frac{2}{N_e}\simeq 0.97 \;,\; r=\frac{12\alpha}{N_e^2}\simeq 0.007
\eeq
as universal observable predictions.
Similar to the simplest fibred inflation models, if the shift by the D3-brane K\"ahler potential was made on $\tau_1$ rather than $\tau_2$, the effective $\alpha=1/2$. The predictions for inflation happening along the radial direction would therefore be
\beq
\alpha = 1/2 \quad \Rightarrow\quad n_s=1-\frac{2}{N_e}\simeq 0.97 \;,\; r=\frac{12\alpha}{N_e^2}\simeq 0.002 \, .
\eeq

These predictions can be altered if the angular directions are active during inflation and truly multifield dynamics takes place (see e.g. \cite{Christodoulidis:2018qdw}). In addition, effects of the WGC precluding semi-infinite plateaus often include steepening from growing corrections~\cite{Pedro:2013pba,Bousso:2013uia,Cicoli:2014bja,Cicoli:2016chb}. We expect these to change the above predictions as well.

In closing the discussion, we wish to note the following: embedding pole N-flation into string theory so far seems to require realizing it in the context of a $K3$ or $T^4$-fibred CY compactification. These models are known to contain another sector capable of driving $\alpha$-attractor inflation~\cite{Kallosh:2017wku} using the two K\"ahler moduli of the fibration, leading to what is known as `fibre inflation'~\cite{Cicoli:2008gp}. The K\"ahler moduli of fibre inflation constitute examples of half-plane fields and contain their own axion partners as the imaginary parts. Applying a WGC based bound in terms of the these half-plane field axions to the field range of fibre inflation itself is a natural question arising from our analysis of pole N-flation, which however falls outside the scope of pole N-flation. Consequently, we leave this   issue for future work.

\section{Conclusions}\label{Sec:Concl}

Pole inflation/$\alpha$-attractors is an intriguing class of models that suggests that the observed primordial power spectrum may be a universal consequence of a pole in the field space metric. That is to say, regardless of a broad range of microphysical considerations, ultimately observables are determined by just a few key parameters characterising the pole. This property is two-sided. On one hand, such a mechanism seriously limits the potential for learning about fundamental physics from cosmology, given there are many fundamental parameters one simply cannot hope to infer from cosmological data. On the other hand, such robust predictions provide an especially appealing target for future observational surveys and in principle would enable a small number of exceptionally sharp statements about the underlying theory. For example, the model predicts that primordial gravitational waves may be detectable. In the context of this model, such a detection would imply the existence of a hyperbolic moduli space~\cite{Kallosh:2015zsa,Carrasco:2015uma}, which in turn may be viewed as indirect evidence for extra dimensions.\footnote{Hyperbolic moduli spaces arise generically in Kaluza-Klein compactification of higher-dimensional Einstein gravity, and hence also in string theory.} To make such statements, however, it is crucial to understand the robustness of the mechanism both from a phenomenological viewpoint and from the perspective of its possible embedding in string theory or another theory of quantum gravity. Considerable progress in this direction has been done by showing that the so-called `fibre inflation' model~\cite{Cicoli:2008gp,Broy:2015zba,Cicoli:2018tcq} is a string realization of $\alpha$-attractors with $\alpha={1/2\,,\,2}$~\cite{Kallosh:2017wku}. Furthermore, investigations on the effects of string moduli backreaction~\cite{Roest:2016lrb} and  K\"ahler corrections~\cite{McDonough:2016der} have given strong evidences of the special resilience of this attractor mechanism. 

In this paper, we have taken a step forward towards a consistent realization of the pole inflation dynamics in string theory, by exploring the possibility of assistance of many fields in the inflaton sector. The proposed {\it pole N-flation} model consists of several open string moduli, such as D3-branes, whose collective motion reduces the distance each brane should traverse in order to yield the inflationary attractor phase. Allowing each individual brane to be sufficiently far from the moduli boundary could improve the radiative stability of this model.

In \S\ref{Sec:WGC}, we focus on the limitations that UV physics imposes on the effective description of pole inflation  when this is embedded into supergravity as a low-energy limit of string theory. We find the existence of axionic partners with decay constants which explicitly depend on the distance to the boundary. This fact has direct consequences for inflation. The bounds which the weak gravity conjecture (WGC) imposes on the periodicity of the axions  ($f\lesssim \Mp$) automatically result in a net constraint on the available length of the exponentially flat plateau typical of pole inflation. We show that in the original single superfield pole-inflation, with a single brane, the inflaton is not even allowed to reach the plateau region of the scalar potential. Moreover, we find that when inflation is driven by the assistance of $N$ branes, these constraints do \emph{not} weaken --- we find that the {\it upper bound} on the canonical radial field range set by the 
WGC scales like 
 in the single field case. 
Rendering the plateau region of the potential available for slow-roll inflation requires relaxing the WGC bound to milder forms.  We interpret these findings as an important   bound on the range of validity of the effective field theory of this cosmological scenario.

The universality of the pole inflation/$\alpha$-attractor mechanism also emerges in our analysis. Despite the presence of $N$ fields, the form of the exponential plateau remains unaltered from the single field case. This implies that when inflation occurs along the collective radial direction, we recover the single field predictions. This may be contrasted with other many-field inflationary constructions, where the predictions at large $N$ are typically distinct from the single field limit \cite{Liddle:1998jc, Kim:2006ys, Kim:2007bc, Easther:2013rva, Price:2014ufa, Dias:2016slx, Hotinli:2017vhx, Bjorkmo:2017nzd} (however, see Ref.~\cite{Bachlechner:2014hsa} for a counter example). That said, a full analysis of the large-$N$ dynamics of this model remains to be explored, as a subset of the angular field directions may also be sufficiently light to play a role in the inflationary dynamics. This may give rise to richer phenomenology through multifield effects which have the capacity to modify the original predictions of the model. While studying the complete dynamics will be a computationally heavy task, the necessary tools have recently been made publicly available \cite{Dias:2015rca, Ronayne:2017qzn, Butchers:2018hds, Dias:2017gva}; we leave this for future work.

Regarding the implementation of pole N-flation in type IIB string theory, while we have made first steps  in  \S\ref{Sec:FibCY} by embedding the model in fibred geometries, developing a consistent program for moduli stabilization within this scenario remains an important step to be addressed. We see this as an exciting avenue to be explored.

\section*{Acknowledgements}

We are grateful to   Shamit Kachru, Renata Kallosh, Andrei Linde,   Eva Silverstein, Yvette Welling and Yusuke Yamada for helpful comments and discussions. We are particularly indebted to Miguel Montero, Jakob Moritz and Irene Valenzuela for numerous illuminating conversations and explanations.  We would also like to acknowledge very useful comments by the JHEP referee.  
AR and AW are grateful to the SITP in Stanford  for   warm hospitality while developing this work. JF, AR and AW are supported by the ERC Consolidator Grant STRINGFLATION under the HORIZON 2020 grant agreement no. 647995. MD is supported by the German Science Foundation (DFG) within the Collaborative Research Centre 676 \textit{Particles, Strings and the Early Universe}. MS is supported by the Research Foundation - Flanders (FWO) and the European Union's Horizon
2020 research and innovation programme under the Marie Sk{\l}odowska-Curie grant agreement No. 665501. MS acknowledges also financial support by `The Foundation Blanceflor Boncompagni Ludovisi n\`ee Bildt', by the `German Academic Exchange Service' (DAAD) and by the foundation `Angelo Della Riccia' for his research stay at DESY.

\section*{Appendix}
\appendix
\section{Properties of angular element}
\label{angles}

The angular element $\Omega_i(\psi_\beta)$ as defined by eq.~\eqref{angularcoords} can be expressed as
\begin{equation}
\Omega_i = \begin{cases}  \cos(\psi_1), & \mbox{if } i= 1 \\ \\ \prod_{\mu=1}^{i-1}\sin(\psi_\mu) \cos(\psi_i) 
, & \mbox{if } 1< i < N \\ \\ \prod_{\mu=1}^{N-1}\sin(\psi_\mu)  , & \mbox{if } i=N \end{cases}
\end{equation}
and therefore
\begin{equation}
\partial_\beta \Omega_i = \begin{cases} 0 , & \mbox{if } i < \beta \\ \\ - \Omega_i \frac{\sin(\psi_\beta)}{\cos(\psi_\beta)} 
, & \mbox{if }   i = \beta  \\ \\ \Omega_i \frac{\cos(\psi_\beta)}{\sin(\psi_\beta)}  , & \mbox{if } i > \beta \end{cases}.
\end{equation}
Using these expressions it is easy to see that, after  take without loss of generality $\beta\geq\gamma$,
\begin{equation}
\sum_{i=1}^N \partial_\beta \Omega_i\partial_\gamma \Omega_i = \sum_{i=\beta}^{N} \partial_\beta \Omega_i\partial_\gamma \Omega_i.
\end{equation}
In the case $\beta \neq \gamma$ it further simplifies
\begin{equation}
\sum_{i=1}^N \partial_\beta \Omega_i\partial_\gamma \Omega_i = \frac{\cos(\psi_\gamma)}{\sin(\psi_\gamma)}\sum_{i=\beta}^{N} \Omega_i \partial_\beta \Omega_i = 0.
\end{equation}
For the case of $\beta=\gamma$, note that 
\begin{equation}
\sum_{i=1}^N (\partial_\beta \Omega_i)^2 = \sum_{i=\beta}^N (\partial_\beta \Omega_i)^2 = \Omega^2_\beta \frac{\sin^2(\psi_\beta)}{\cos^2(\psi_\beta)} + \sum^{N}_{i=\beta+1}\Omega^2_i \frac{\cos^2(\psi_\beta)}{\sin^2(\psi_\beta)} 
\label{andereq}
\end{equation}
which is in general not zero. For example, when $\beta=1$ this reduces to 1:
\begin{equation}
\sum_{i=1}^N (\partial_\beta \Omega_i)^2 = \frac{1}{2} \partial_{\beta\beta}^2  \sum_{i=1}^N \Omega_i^2 -  \sum_{i=1}^N \Omega_i \partial_{\beta\beta}^2 \Omega_i = - \sum_{i=1}^N \Omega_i \partial_{\beta\beta} \Omega_i \, = 1.
\end{equation}
\\

Throughout the paper it is of special interest  the configuration where all $\Omega_i$ are the same and therefore $\Omega_i^2=1/N$. In order to derive   eq.~\eqref{psi_ev}, note that 
\begin{align}
&\Omega^2_{1}=\cos^2(\psi_1)=\frac{1}{N} \	\	\ \rightarrow  \		\	\	\frac{\cos^2(\psi_1)}{\sin^2(\psi_1)}=\frac{1}{N-1} \\ \nonumber
&\Omega^2_{2}=\sin^2(\psi_1)\cos^2(\psi_2)=\frac{1}{N}  \	\	\ \rightarrow  \		\	\	\cos^2(\psi_2)=\frac{1}{N-1}  \	\	\ \rightarrow  \		\	\	\frac{\cos^2(\psi_2)}{\sin^2(\psi_2)}=\frac{1}{N-2} \\ \nonumber
\vdots \\ \nonumber 
&\frac{\cos^2(\psi_\beta)}{\sin^2(\psi_\beta)} = \frac{1}{N-\beta} \ .
\end{align} 
This relation, together with   eq.~\eqref{andereq}, implies that in the configuration of interest
\begin{equation}
\sum_{i=1}^N (\partial_\beta \Omega_i)^2 = \frac{N-\beta+1}{N} \ .
\end{equation}

\bibliographystyle{JHEP}

\bibliography{references}

\end{document}